\documentclass[reprint,aps,showpacs,preprintnumbers,amsmath,amssymb,superscriptaddress]{revtex4-1}

\usepackage[pdftex]{graphicx}
\DeclareGraphicsExtensions{.pdf}
\usepackage{dcolumn}
\usepackage{bm}
\usepackage{wasysym}
\usepackage{xspace}
\usepackage{amssymb}
\usepackage[mediumqspace, thinspace, amssymb]{SIunits}
\usepackage{color}

\newcommand{\be}{\begin{equation}}
\newcommand{\ee}{\end{equation}}

\newcommand{\bea}{\begin{eqnarray}}
\newcommand{\eea}{\end{eqnarray}}

\newcommand{\fst}{\quad\textrm{.}}%
\newcommand{\com}{\quad\textrm{,}}%

\newcommand{\3}{(0~$\overline{3}$~0)\xspace}
\newcommand{\1}{(0~$\overline{1}$~0)\xspace}
\newcommand{\PCMO}{Pr$_{0.5}$Ca$_{0.5}$MnO$_3$\xspace}

\begin{document}

\title{Disentangling charge and structural contributions during coherent atomic motions studied by ultrafast resonant x-ray diffraction}

\author{L. Rettig}
\email[Corresponding author: ]{rettig@fhi-berlin.mpg.de}
\affiliation{Swiss Light Source, Paul Scherrer Institut, 5232 Villigen PSI, Switzerland}
\affiliation{Abteilung Physikalische Chemie, Fritz-Haber-Institut der Max-Planck-Gesellschaft, Faradayweg 4-6, D-14195 Berlin, Germany}
\author{A. Caviezel}
\affiliation{Swiss Light Source, Paul Scherrer Institut, 5232 Villigen PSI, Switzerland}
\author{S. O. Mariager}
\affiliation{Swiss Light Source, Paul Scherrer Institut, 5232 Villigen PSI, Switzerland}
\author{G. Ingold}
\affiliation{Swiss Light Source, Paul Scherrer Institut, 5232 Villigen PSI, Switzerland}
\affiliation{SwissFEL, Paul Scherrer Institut, 5232 Villigen PSI, Switzerland}
\author{C. Dornes}
\affiliation{Institute for Quantum Electronics, ETH Z\"urich, 8093 Z\"urich, Switzerland}
\author{S-W. Huang}
\affiliation{Swiss Light Source, Paul Scherrer Institut, 5232 Villigen PSI, Switzerland}
\author{J. A. Johnson}
\affiliation{Swiss Light Source, Paul Scherrer Institut, 5232 Villigen PSI, Switzerland}
\author{M. Radovic}
\affiliation{Swiss Light Source, Paul Scherrer Institut, 5232 Villigen PSI, Switzerland}
\affiliation{SwissFEL, Paul Scherrer Institut, 5232 Villigen PSI, Switzerland}
\author{T. Huber}
\affiliation{Institute for Quantum Electronics, ETH Z\"urich, 8093 Z\"urich, Switzerland}
\author{T. Kubacka}
\affiliation{Institute for Quantum Electronics, ETH Z\"urich, 8093 Z\"urich, Switzerland}
\author{A. Ferrer}
\affiliation{Swiss Light Source, Paul Scherrer Institut, 5232 Villigen PSI, Switzerland}
\affiliation{Institute for Quantum Electronics, ETH Z\"urich, 8093 Z\"urich, Switzerland}
\author{H. T. Lemke}
\altaffiliation[current address: ]{SwissFEL, Paul Scherrer Institut, 5232 Villigen PSI, Switzerland}
\affiliation{LCLS, SLAC National Accelerator Laboratory, Menlo Park, California 94025, USA}
\author{M. Chollet}
\affiliation{LCLS, SLAC National Accelerator Laboratory, Menlo Park, California 94025, USA}
\author{D. Zhu}
\affiliation{LCLS, SLAC National Accelerator Laboratory, Menlo Park, California 94025, USA}
\author{J. M. Glownia}
\affiliation{LCLS, SLAC National Accelerator Laboratory, Menlo Park, California 94025, USA}
\author{M. Sikorski}
\affiliation{LCLS, SLAC National Accelerator Laboratory, Menlo Park, California 94025, USA}
\author{A. Robert}
\affiliation{LCLS, SLAC National Accelerator Laboratory, Menlo Park, California 94025, USA}
\author{M. Nakamura}
\altaffiliation[current address: ]{PRESTO, Japan Science and Technology Agency (JST), Kawaguchi, 332-0012, Japan}
\affiliation{RIKEN Center for Emergent Matter Science (CEMS),Wako 351-0198, Japan}
\author{M. Kawasaki}
\affiliation{Department of Applied Physics and Quantum-Phase Electronics Center, University of Tokyo, Hongo, Tokyo 113-8656, Japan}
\affiliation{RIKEN Center for Emergent Matter Science (CEMS),Wako 351-0198, Japan}
\author{Y. Tokura}
\affiliation{Department of Applied Physics and Quantum-Phase Electronics Center, University of Tokyo, Hongo, Tokyo 113-8656, Japan}
\affiliation{RIKEN Center for Emergent Matter Science (CEMS),Wako 351-0198, Japan}
\author{S. L. Johnson}
\affiliation{Institute for Quantum Electronics, ETH Z\"urich, 8093 Z\"urich, Switzerland}
\author{P. Beaud}
\affiliation{Swiss Light Source, Paul Scherrer Institut, 5232 Villigen PSI, Switzerland}
\affiliation{SwissFEL, Paul Scherrer Institut, 5232 Villigen PSI, Switzerland}
\author{U. Staub}
\email{urs.staub@psi.ch}
\affiliation{Swiss Light Source, Paul Scherrer Institut, 5232 Villigen PSI, Switzerland}

\date{\today}

\begin{abstract}
We report on the ultrafast dynamics of charge order and structural response during the photoinduced suppression of charge and orbital order in a mixed-valence manganite. Employing femtosecond time-resolved resonant x-ray diffraction below and at the Mn K absorption edge, we present a method to disentangle the transient charge order and structural dynamics in thin films of \PCMO. Based on the static resonant scattering spectra, we extract the dispersion correction of charge ordered Mn$^{3+}$ and Mn$^{4+}$ ions, allowing us to separate the transient contributions of purely charge order from structural contributions to the scattering amplitude after optical excitation. Our finding of a coherent structural mode at around $\unit{2.3}{THz}$, which primarily modulates the lattice, but does not strongly affect the charge order, confirms the picture of the charge order being the driving force of the combined charge, orbital and structural transition.

\end{abstract}

\pacs{64.60.A-, 61.05.cp, 71.27.+a, 78.47.J-}

\maketitle

\section{Introduction}
The coupling between the crystal and the electronic structure is of great importance for the physical properties of materials. A particular interest lies in materials, which have strong correlation between the electronic, orbital, magnetic and structural degrees of freedom. These interactions often can lead to new ground states of the materials, which are characterized by induced orders in one or various of the subsystems, such as e.g. superconducting states, structural phase transitions or charge-density waves. Often, several orders coexist in a material, and their cooperative or competing character is fundamental for the material properties. In particular, it is of great interest to identify the primary instability that drives the phase transition, and a possible parasitic order. One example for such a state is the structural-nematic phase transition in the Fe pnictide parent compounds, where electronic nematic order is considered to drive a concomitant structural phase transition~\cite{Fernandes2014}.

To understand these couplings, a promising route is to manipulate e.g. the crystal structure and investigate the corresponding changes of the other degrees of freedom. A widely used way is to use epitaxial strain to modify the crystal structure or to apply high pressure. Though these approaches are very useful and have led to very interesting observations, they both are rather limited in their applicability and might change also the microstructure of the materials. In addition, strongly coupled phase transitions show in equilibrium often identical behavior as function of temperature or pressure, and understanding their hierarchy is therefore very challenging.

These limitations can be overcome by studying the material response to an ultrafast stimulus, which is faster than the fundamental interaction timescale between the degrees of freedom. One very powerful way is to excite the structure directly by THz or mid-infrared radiation and test the ultrafast electronic, magnetic or structural responses to the excitation. This approach has led to very interesting variation of the electronic or magnetic properties of materials in several correlated electron systems~\cite{Rini2007, Forst2011a, Forst2013, Fausti2011, Kubacka2014, Esposito2017, Mitrano2016}. Structural modifications induced by such excitations have also been directly investigated by ultrafast x-ray diffraction~\cite{Gruebel2016, Mankowsky2014, Kozina2017}.

Another approach, though less direct, is to electronically excite a material by an optical excitation and determine the structural and electronic changes in the time domain. A particular interesting case is when a coherent phonon is created by very short optical excitations. By using selective ultrafast probes that are directly sensitive to an ultrafast response of a specific subsystems to such an excitation, one can gain fundamental insight into the underlying processes. For instance, by using ultra short x-ray pulses, one can follow the structural motion, and the actual size of the corresponding distortion can be quantified~\cite{Johnson2009}. By additionally detecting the changes of the electronic structure e.g. by photoemission, one can obtain information on the mode-dependent electron phonon coupling, as has been recently demonstrated for several Fe-pnictide and chalcogenide compounds~\cite{Rettig2015, Gerber2015, Gerber2017}. In addition, studying the temporal evolution of the structural phase transition with the transient nematic order provided strong evidence for nematic ordering as the driving force of the transition~\cite{Rettig2016}.

Manganites are another class of interesting materials, in which the correlation between electronic ordering, and the associated structural distortions have been intensively studied in the time domain~\cite{Ogasawara2001, Polli2007, Beaud2009, Beaud2014, Singla2013, Esposito2017,Esposito2018}. The half doped simple perovskite Pr$_{0.5}$Ca$_{0.5}$MnO$_3$ (PCMO) shows a charge and orbital order (CO/OO) phase transition around $\unit{240}{K}$ concomitant with a structural distortion, characterized by an alternating pattern of Mn$^{3+}$ and Mn$^{4+}$ ions, schematically shown in Fig. 1(b). This transition is followed in temperature by an antiferromagnetic order of the Mn$^{3+}$ spins around $\unit{150}{K}$~\cite{Zhou2011}. The correlation between electronic and crystal structure has been studied by ultrafast time-resolved resonant and non-resonant x-ray diffraction~\cite{Beaud2014}. In particular, reflections that are selectively sensitive to electronic order, the orbital order or the structural distortions have been investigated. The coherent dynamics of all of these reflections and the optically driven phase transition could be well described by an ultrafast quench of the charge order with a time-dependent order parameter that triggers the structural phase transition and launches several coupled coherent phonon modes~\cite{Beaud2014}. 

However, an important open question remained whether the coherent modes also couple back onto the charge order and coherently modulate the charge disproportionation between neighboring Mn sites. Although the mode amplitude for the \3 reflection, which is primarily sensitive to the electronic ordering, has been found to be much weaker as compared to that of reflections sensitive to the structural distortion alone, a clear separation of electronic and structural components was beyond the scope of that study. To clarify this point and to better understand the hierarchy of coupled phase transitions in \PCMO, we address here how the electronic states are modified during the coherent motion of the ions involved in the long lived low frequency coherent phonon oscillation. This is achieved by disentangling the electronic charge order dynamics and the structural components using on- and off-resonant time-resolved diffraction data. This allows us to determine the role of this mode in relation to the electronic ordering.

\begin{figure}[tb]
\includegraphics[width=8.6cm]{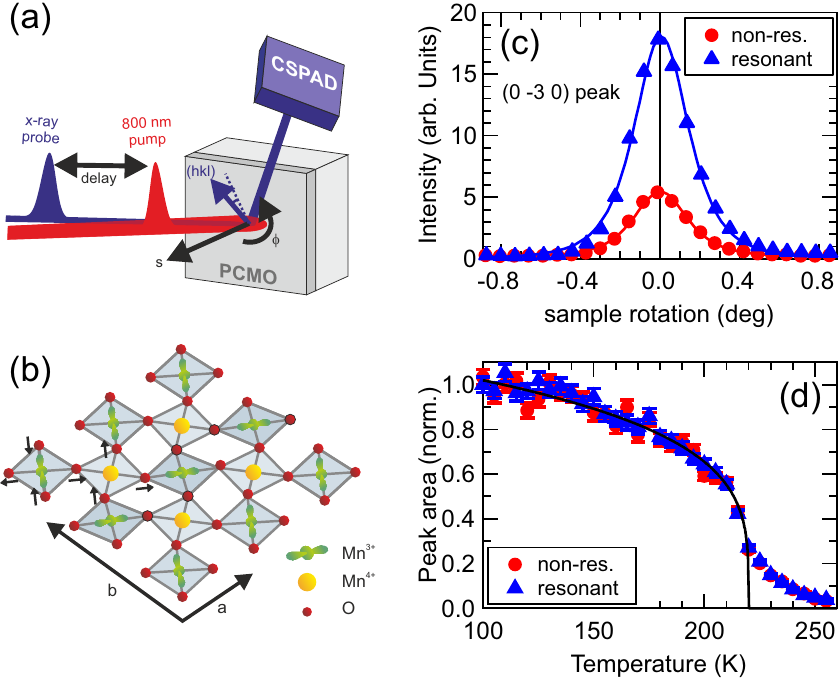}  
\caption{
\label{fig:fig1}
Experimental scheme and static x-ray diffraction results. (a) Schematic of the time-resolved resonant x-ray diffraction experiment. (b) Sketch of the charge and orbitally ordered phase of alternating Mn$^{3+}$ and Mn$^{4+}$ ions in the a/b plane. (c) Static rocking curves of the \3 reflection measured by a sample rotation about the surface normal, below (red circles, $\unit{6.530}{keV}$) and at resonance (blue triangles, $\unit{6.555}{keV}$) with the Mn K-absorption edge. (c) Temperature dependence of the normalized integrated on- and off-resonant peak intensity. The solid black line is a power law fit. In good agreement with transport measurements we find $T_\mathrm{COO}\approx\unit{220}{K}$.
}
\end{figure}

\section{Experimental Details}

Resonant x-ray diffraction experiments were performed on a thin film of \PCMO of approximately $\unit{40}{nm}$ thickness grown on (011)-oriented (LaAlO$_3$)$_{0.3}$ - (SrAl$_{0.5}$Ta$_{0.5}$O$_3$)$_{0.7}$ (LSAT) substrate using a pulsed laser deposition technique with a laser pulse frequency of $\unit{2}{Hz}$ at $\unit{850^\circ}{C}$ in an oxygen pressure of 1.5 mTorr (for details see Ref. \citep{Okuyama2009}). 

Static energy-dependent x-ray diffraction experiments were performed at the X04SA beam line at the Swiss Light Source, Paul Scherrer Institute using the surface science scattering end station~\cite{Willmott2013} equipped with a Pilatus 200k pixel detector~\cite{Broennimann2006}, and time-resolved on- and off-resonant diffraction experiments were performed at the X-ray Pump-Probe (XPP) instrument~\cite{Chollet2015} at the Linac Coherent Light Source (LCLS)~\cite{White2015} at the SLAC National Accelerator Laboratory. In both experiments, an asymmetric diffraction configuration~\cite{Johnson2008} as sketched in Fig.~\ref{fig:fig1}(a) was used, where the horizontally polarized x-ray probe pulses entered the film at $8^\circ$ grazing incidence. The x-ray energy was tuned in the vicinity of the Mn K edge using a silicon (111) monochromator, with an energy resolution of $\approx\unit{1.1}{eV}$. During energy scans, the sample was held in diffraction condition (constant $\mathbf{q}$ scans). The sample temperature was controlled between 100 K and the charge ordering temperature of $T_{COO}\approx\unit{220}{K}$ using a cryogenic nitrogen blower.

For time-resolved x-ray diffraction experiments performed at the LCLS, a weakly focused ($\unit{230 \times 230}{\mu m^2}$) $\unit{55}{fs}$ optical pulse with a wavelength of $\unit{800}{nm}$ excited the sample with a repetition rate of $\unit{120}{Hz}$ at an incidence angle of $14^\circ$ (p-polarization), synchronized to the $\approx\unit{50}{fs}$ x-ray probe pulses from the LCLS, and diffracted x-ray pulses were detected at the full repetition rate using the Cornell-SLAC Pixel Array Detector (CSPAD)~\cite{Herrmann2013}. The timing jitter between pump and probe pulses was measured and corrected shot-by-shot using the spectral encoding correlation technique\cite{Harmand2013}, with an accuracy of $\approx\unit{15}{fs}$, yielding an overall temporal resolution of $\approx\unit{85}{fs}$.

\section{Results and discussion}

The charge ordering manifests in a lowering of crystal symmetry that is accompanied by a structural distortion and a staggered configuration of the Mn$^{3+}$ orbitals. This results in the occurrence of additional, symmetry forbidden reflections in the x-ray diffraction signal. The diffraction intensity of some of these reflections becomes strongly modulated in the vicinity of the Mn K-absorption edge due to the charge disproportionation of neighboring Mn sites. Fig.~\ref{fig:fig1}(c) shows x-ray rocking curves of the high-temperature symmetry-forbidden \3 reflection, measured by rotating the sample about the surface normal, below (red) and at the Mn K-absorption edge (blue). Here, the strong resonant enhancement of the intensity by $> \times 3$ at resonance arises due to the charge ordering pattern, while the intensity off-resonance originates from the accompanying structural distortion. The temperature dependence of the integrated peak intensity of this reflection below and at resonance is shown in Fig.~\ref{fig:fig1}(d), which represents a regular behavior of a reflection that follows the order parameter squared. The equivalent temperature dependence of the on- and off-resonant diffraction intensity demonstrates the strong coupling of the structural and charge order transitions in thermal equilibrium in contrast to results found in nickelates~\cite{Staub2002}. 

\subsection{Refinement of static resonant x-ray diffraction data}

\begin{figure}[tb]
\includegraphics[width=8.6cm]{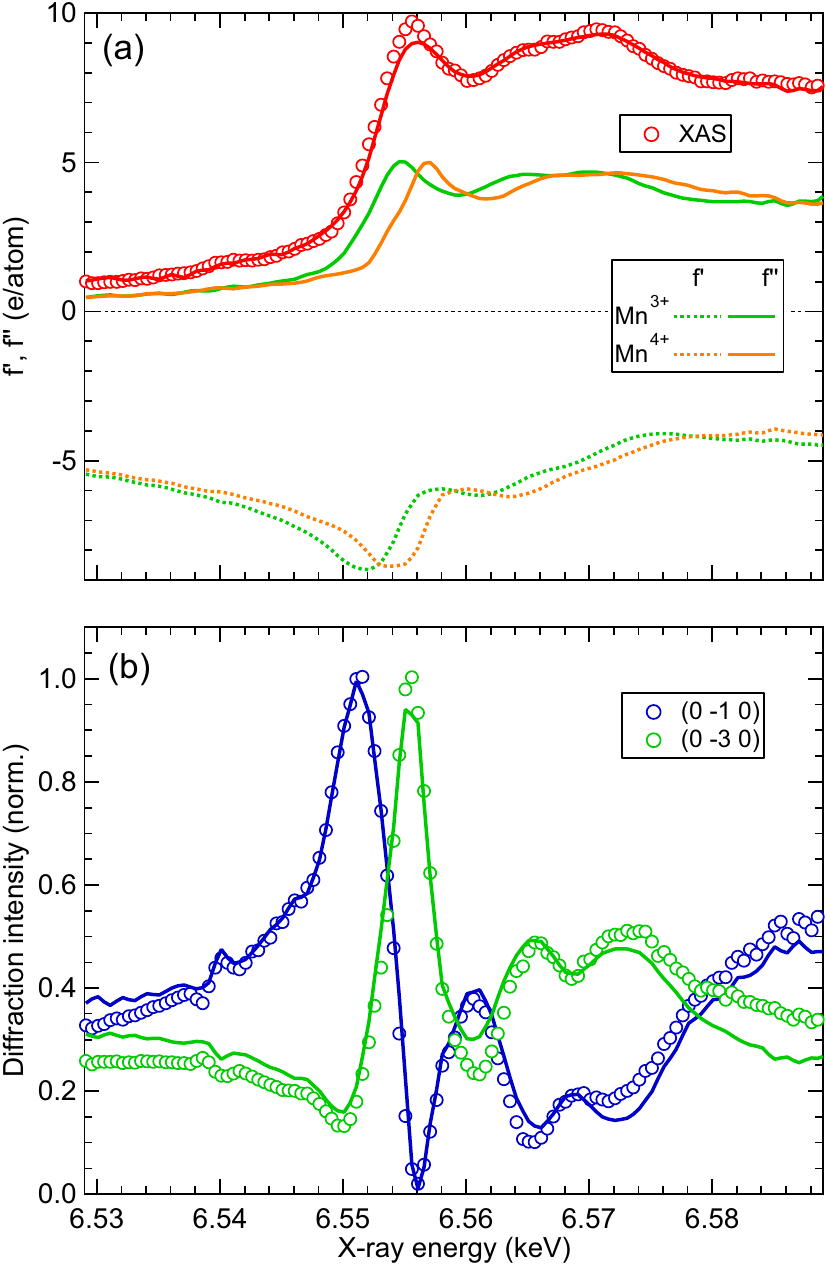}  
\caption{
\label{fig:fig2}
Energy dependence of resonant x-ray diffraction. (a) X-ray fluorescence signal (red) and (b) resonant diffraction intensity of the \1 (blue) and \3 (green) peaks across the Mn K-absorption edge. The solid red, blue and green lines are the results of a refinement of the data (see text). The extracted real (imaginary) part of the anomalous dispersion corrections of Mn$^{3+}$ and Mn$^{4+}$ sites is shown in (a) as green and yellow dashed (solid) lines, respectively.
}
\end{figure}

The resonant enhancement of the x-ray diffraction signal of a charge order reflection arises due to the valence charge inequality of neighboring ions, which gives rise to a shift of the absorption edge, and therefore to a resonant contribution from the x-ray dispersion correction close to an absorption edge. As the film is very thin compared to the x-ray penetration depth, an absorption correction, which typically creates large uncertainties on such data of single crystals, is negligible in our study. In addition, as the peaks are broad due to the finite size of the film, also refraction effects can be ignored when changing the x-ray energy. Note that we consider here only the electronic difference (labeled as charge order/disproportionation) on the Mn sites as seen by the dipole transition at the Mn K-edge that probes the Mn 4p states. This analysis remains valid independently of the microscopic origin of the resonant behavior e.g. due to orbital contributions or phenomena such as bond valence order~\cite{Garcia2001,Zimmermann2001, Zimmermann2001a, Garcia2003, Zimmermann2003, Grenier2004, Herrero-Martin2012}. Distinguishing such models would require non trivial first principle calculations of the spectral shape of the scattering factors at resonance that go beyond to goal of this study. The charge disproportionation and the respective dispersion corrections in such an electronic ordered system can directly be determined from resonant x-ray diffraction data of selected symmetry forbidden charge order peaks of type (0 $k$ 0) with $k$ odd~\cite{Murakami1998,Staub2002, Mulders2009}. For the case of \PCMO, their resonant structure factor near the Mn K-edge can be written to first order as
\begin{widetext}
\be
F_{(0k0)} = \underbrace{A_{\textrm{Pr/Ca,O}}(Q, E)}_{\mathrm{structure} ~ \sigma} +\underbrace{4\Delta f^0_\mathrm{Mn}(Q) + 4\Delta f'_\mathrm{Mn}(E) + 4i \Delta f''_\mathrm{Mn}(E)}_{\mathrm{charge ~ order} ~ \eta}\fst
\ee
\end{widetext}
Here, the first term corresponds to the structural contributions from the displacements of Pr, Ca and O ions. The contribution from the charge disproportionation is determined by the form factor differences $\Delta f_\mathrm{Mn} = f_\mathrm{Mn^{3+}} - f_\mathrm{Mn^{4+}}$, and consists of three parts, which represent the change in Thompson scattering amplitude and the difference in the real and imaginary part of the dispersion correction between the charge ordered ions, respectively. With knowledge of the low-temperature structure~\cite{Rodriguez2005} this expression for the structure factor can be used to describe the resonant diffraction intensity $I\propto|F(Q,E)\cdot F^*(Q,E)|$, and to determine the dispersion corrections $f_\mathrm{Mn^{3+}}$ and $f_\mathrm{Mn^{4+}}$. Additionally, according to the optical theorem the x-ray absorption spectrum is proportional to the average imaginary part of the dispersion correction $f^{''}_\mathrm{Mn^{3+}} + f^{''}_\mathrm{Mn^{4+}}$, which is fitted simultaneously. 

We adopt the following iterative global fit procedure to determine the dispersion corrections $f_\mathrm{Mn^{3+}}$ and $f_\mathrm{Mn^{4+}}$ from the x-ray fluorescence spectrum (Fig.~\ref{fig:fig2}(a)) and the resonant x-ray diffraction spectra for the \1 and \3 reflections (Fig.~\ref{fig:fig2}(b)): In each iteration, the imaginary parts of the dispersion corrections $f^{''}_\mathrm{Mn^{3+}} + f^{''}_\mathrm{Mn^{4+}}$ are varied, and the real parts $f^{'}_\mathrm{Mn^{3+}} + f^{'}_\mathrm{Mn^{4+}}$ are determined by Kramers-Kronig transformation. From these, the resonant diffraction intensities for the two peaks are calculated, and together with the absorption spectrum fitted to the experimental data. Additionally, a regularization term of the form
\be
\lambda \left(f^{''}_\mathrm{Mn^{3+}}(E)-f^{''}_\mathrm{Mn^{4+}}(E+\delta E)\right)^2\com
\ee 
that favors resemblance of the individual $f^{''}$ curves with an energy shift $\delta E$ is introduced, where $\lambda$ controls the importance of this condition.

Results of this iterative fitting procedure are shown as solid lines in Fig.~\ref{fig:fig2} and reproduce all data reasonably well. In particular, the shifted maxima of the resonant diffraction curves are well captured by the model. Additionally, Fig.~\ref{fig:fig2}(a) shows the determined real and imaginary parts of the dispersion correction for Mn$^{3+}$ and Mn$^{4+}$. The fit yields an energy shift of the absorption edge of $\delta E\approx\unit{2.2}{eV}$, which in comparison to the single-valence compounds CaMnO$_3$ and LaMnO$_3$ would correspond to a valence state of Mn$^{3.2+}$ and Mn$^{3.8+}$ in the approximation of spherical scattering factors~\cite{Garcia2001}. These extracted dispersion corrections can now be used to disentangle the time-dependence of structural and charge order dynamics from measurements taken at two different energies.

\subsection{Time-dependent structural and charge order dynamics}

\begin{figure}[tb]
\includegraphics[width=8.6cm]{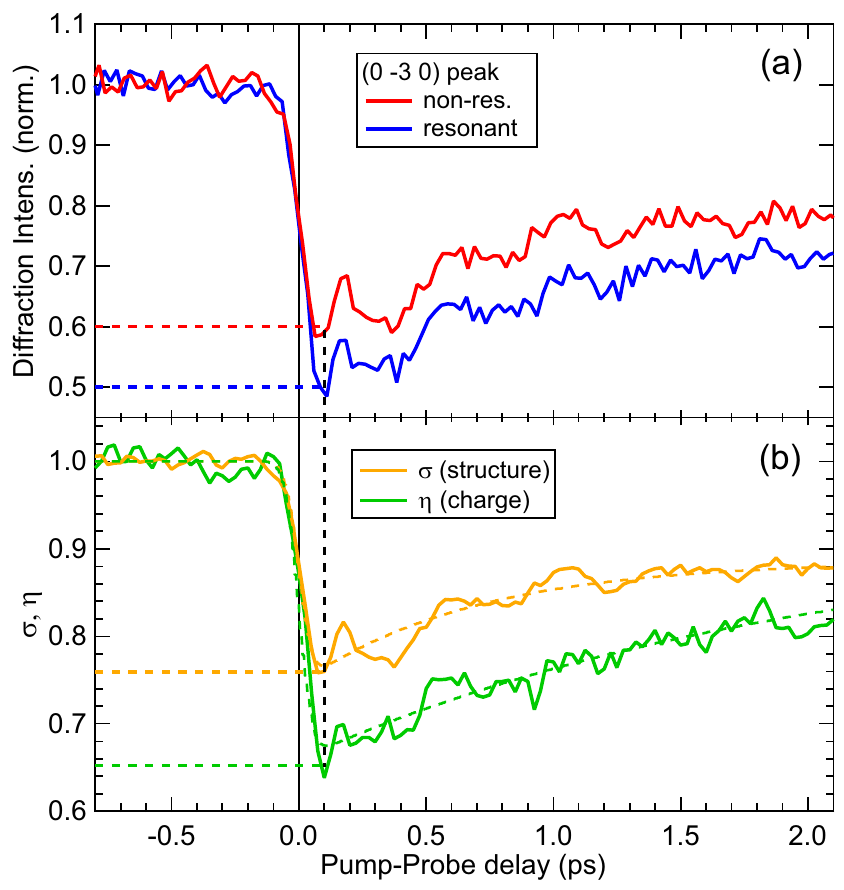}  
\caption{
\label{fig:fig3}
Time-dependent resonant x-ray diffraction. (a) Normalized time-dependent  diffraction intensity of the \3 peak measured off-resonant (red, $\unit{6.530}{keV}$) and resonant (blue, $\unit{6.550}{keV}$) with the Mn K-absorption edge, respectively. (b) Extracted time-dependent structural ($\sigma$, yellow) and charge ($\eta$, green) order parameters. Dashed lines are single-exponential fits (see text).
}
\end{figure}

Fig.~\ref{fig:fig3}(a) shows the time-resolved diffraction intensities of the \3 reflection at two different x-ray energies below resonance (red) and at the maximum of the resonant intensity (blue), with an incident excitation fluence of $F=\unit{1.8}{mJ/cm^2}$. Both curves show a fast drop within the time-resolution of the experiment, followed by a weak recovery on a picosecond timescale, overlaid by weak coherent oscillations. Apart from a slightly larger suppression at resonance, the two curves look very similar.

\begin{figure}[tb]
\includegraphics[width=8.6cm]{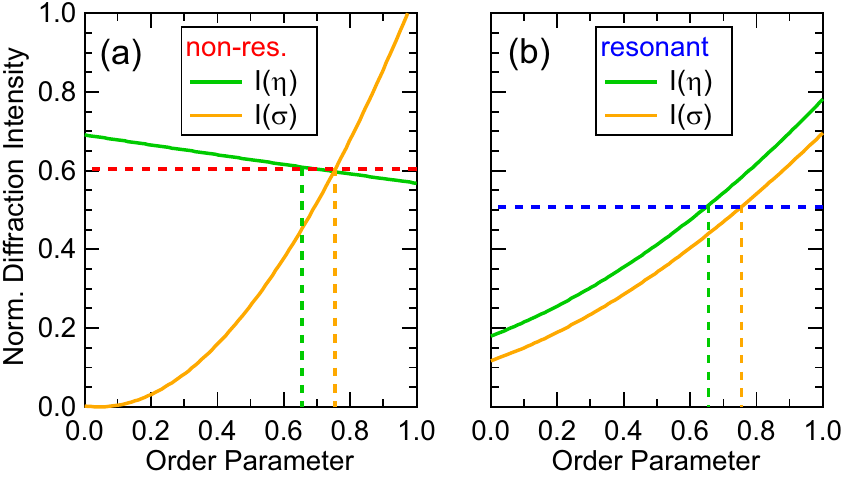}  
\caption{
\label{fig:fig4}
Determination of the transient order parameters. Dependence of the normalized diffraction intensity on the structural (yellow) and charge order parameters (green) (a) off-resonant and (b) on resonance with the Mn K-absorption edge. Dashed lines mark the values of the order parameters obtained at the minimum of the diffraction intensity time traces. The respective other order parameter is kept fix at this value.
}
\end{figure}

In order to separate the dynamics of charge order and structural distortion, we rewrite equation (1) introducing time-dependent structural and charge order parameters $\sigma (t)$ and $\eta (t)$, respectively:
\be
F^E_{(0\overline{3}0)}(t) = \sigma(t) \cdot C^E_1 + \eta(t) \cdot C^E_2 \com
\ee
with $C^E_1 = A_{\textrm{Pr/Ca,O}}(E)$ and $C^E_2 = 4\Delta f^0_\mathrm{Mn} + 4\Delta f'_\mathrm{Mn}(E) + 4i \Delta f''_\mathrm{Mn}(E)$, and $E=\unit{6.530}{keV}$ and $\unit{6.555}{keV}$. This expression allows us to investigate the sensitivity of the diffraction intensity to the structural distortion and to the charge ordering. The corresponding diffraction intensities normalized to the value at $\sigma=\eta=1$ are shown in Fig.~\ref{fig:fig4}(a) and (b) as a function of the size of the corresponding order parameters for the two investigated energies, respectively. Here, green curves show the dependence on $\eta$, while yellow curves show the dependence on $\sigma$, and the respective other order parameter is held at the value indicated by the dashed green and yellow lines. Due to the fact that the charge ordering mostly influences the diffraction signal at resonance, the off-resonant diffraction intensity is, as expected, mostly sensitive to the structural distortion, and only shows a weak variation with the charge order parameter $\eta$. In contrast, at the resonance energy, both structural and charge order parameters show a similar influence on the diffraction intensity, demonstrating the necessity of a proper description of the energy-dependent x-ray intensity to disentangle charge and structural order dynamics.

This is achieved by inverting the expressions for the normalized diffraction intensities 
\be
\left(\frac{I(t)}{I_0}\right)^E_{(0\overline{3}0)} = \left|\left(\frac{F(t)}{F_0}\right)^E_{(0\overline{3}0)}\right|^2
\ee
evaluated at the two measured energies to yield the corresponding structural and charge order parameters $\sigma(t)$ and $\eta(t)$, respectively. The disentangled normalized time-dependent structural and charge order parameters are shown in Fig.~\ref{fig:fig3}(b). Both order parameters show a very fast drop within the temporal resolution of $\sim\unit{80}{fs}$, followed by a slow recovery on a picosecond timescale. The charge order parameter $\eta (t)$ shows a significantly stronger suppression, with a minimum of $\sim0.65$ at $\unit{100}{fs}$, than the structural order parameter $\sigma (t)$, which shows a minimum of $\sim0.76$ at the same time. As a crosscheck, we can calculate the diffraction intensities for the two energies from our diffraction model for this time delay (Fig.~\ref{fig:fig4}). The values of $\sim0.6$ for the non-resonant and $\sim0.5$ for the resonant normalized diffraction intensity agree well with the observed intensities at $t=\unit{100}{fs}$, in particular their ratio matches very well, demonstrating the consistency of our evaluation.

\begin{figure}[tb]
\includegraphics[width=8.6cm]{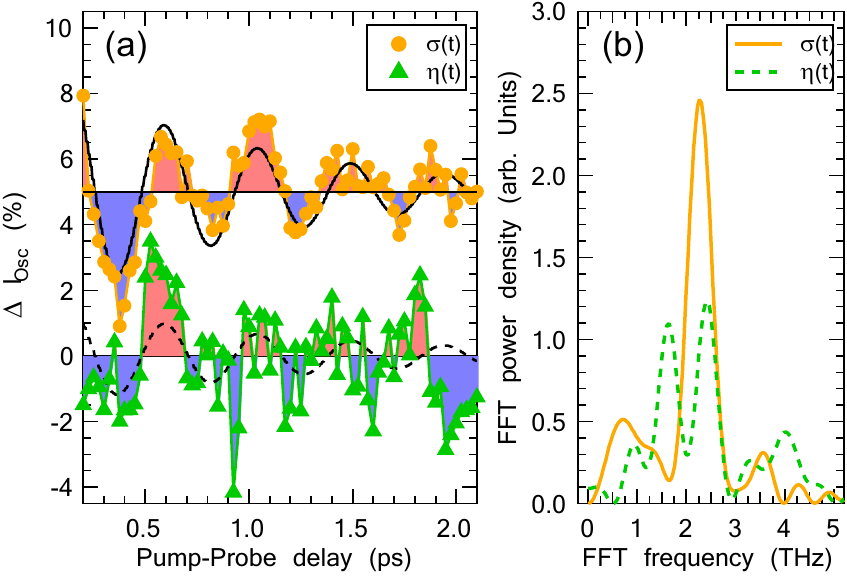}  
\caption{
\label{fig:fig5}
Coherent oscillations of structural and charge components. (a) Oscillatory component of the structural (yellow) and the charge order parameter (green) after subtraction of the single-exponential background shown in fig~\ref{fig:fig3}. Data are vertically offset, and lines are damped cosine fits (see text). (b) Fast Fourier transform power density of the data shown in (a). 
}
\end{figure}

The stronger response of the charge order parameter suggests the suppression of charge order as the driving force for the structural dynamics, in agreement with our previous description of the coupled charge, orbital and structural dynamics by a time-dependent charge order parameter~\cite{Beaud2014}. However, how strongly the structural dynamics could also influence the charge order dynamics on the ultrafast timescale is still an open question. Apart from the strong suppression around $t_0$, weak coherent oscillations from coherently excited phonon modes are visible in the raw diffraction data. Interestingly, however, these oscillations predominantly show up in the structural order parameters. For a more quantitative analysis, a smooth exponential background function is subtracted from the curves in Fig.~\ref{fig:fig3}(b), and the residual coherent signal is shown in Fig.~\ref{fig:fig5}(a) for the two order parameters. Fits with a damped sinusoidal function yield a reasonable description (adjusted $R^2=0.66$) of $\sigma(t)$ with an oscillation amplitude of $2.8\%$ at a frequency of $\unit{2.2}{THz}$ (black solid line), while a similar fit of $\eta(t)$ cannot describe the data well (dashed black line, adjusted $R^2=0.07$). This is corroborated by the Fast Fourier transform (FFT) of the two order parameters shown in Fig.~\ref{fig:fig5}(b). Here, $\sigma(t)$ shows a strong spectral peak at $\sim\unit{2.3}{THz}$, very close to the dominating coherent phonon frequency of in-plane motion of the Pr/Ca ions~\cite{Beaud2014}. In contrast, the spectrum of $\eta(t)$ shows only weak features near $\unit{1.5}{THz}$ and $\unit{2.5}{THz}$, which are close to the noise limit. Therefore we conclude, that the coherent phonon mode that governs the structural transition does not couple back on the charge order parameter.

The observed coherent mode around $\unit{2.3}{THz}$ agrees well with the coherent modes observed in \PCMO and similar manganites~\cite{Beaud2009,Beaud2014,Caviezel2012,Caviezel2013, Lim2005}, which have been identified as coherent oscillations of the Pr and Ca cations coupled to lateral motions of the Mn-O cages~\cite{Caviezel2013}, driven by the suppression of charge order and Jahn-Teller distortion~\cite{Beaud2014}. This mode does not soften significantly for increasing temperatures when approaching the charge and orbital ordering transition~\cite{Caviezel2012}, suggesting that this mode is not a true soft mode of the transition and is therefore not relevant for the electronic ordering. The amplitude mode of the charge order is expected to be linked to a change of the volume of the oxygen octahedra as a change in ionic charge impacts the ionic radii of the Mn ions. Consequently, a difference in charge does expand and shrink the size of the octahedra alternatingly along the ordering wave vector, as has been observed by the structural determination in the ground state~\cite{Rodriguez2005}. Candidates for the relevant modes for the charge and orbital order are the asymmetric stretch and Jahn-Teller modes of the distorted octahedra, respectively, which occur at much faster frequencies of around $\unit{16}{THz}$~\cite{Singla2013,Beaud2014}. The initial dynamics of those oscillations occur within the first $\sim\unit{100}{fs}$ after excitation and are not accessible with our current temporal resolution. However, due to the strongly coupled nature of the various modes in the system, the slow coherent cation oscillation dominates the dynamics at later times also of those faster modes, and leads to a coherent modulation of their coordinates~\cite{Beaud2014}. In an intuitive picture, this can be seen as a harmonic oscillator driven off-resonantly, which also oscillates with the driving frequency. This implies that the coherent oscillations found in the structural dynamics indeed correspond to a modulation of the octahedra volume along the charge order coordinate. Our observation of absence of a coherent response in the charge order itself thus demonstrates that those coherent structural modulations of the octahedral volumes do not lead to a transient charge transfer between neighboring Mn sites. 

This brings us back to our initial question about the hierarchy of coupled phase transitions in this system. The absence of charge order response to a structural modulation demonstrates that the charge and orbital order is the driving force of the coupled phase transition, and the structural distortion can be regarded as a secondary order, which stabilizes the electronic order but is not sufficient to drive the transition alone. This is an analogous behavior to the coupled structural/nematic transition in the Fe pnictide parent compounds, where also strong evidence for an electronically driven phase transition and a secondary structural transition exists~\cite{Fernandes2014,Rettig2016}. The identification of the hierarchy of coupled phase transitions in complex materials as demonstrated in our experiment does not only provide an important benchmark for theory, but could also allow identifying the appropriate handle for precise control of phase transitions with competing orders, such as the charge-density wave or stripe order ground state present in superconducting cuprates.

\section{Conclusion}
In conclusion, we investigated the transient coupling between charge order and structural response in charge and orbitally ordered \PCMO. Using time-resolved x-ray diffraction both below and at the Mn K absorption edge, we were able to separate contributions of purely charge order from structural contributions to the scattering amplitude. We find a coherent structural mode, which primarily modulates the lattice, but does not strongly affect the charge order. Our findings confirm the charge order to be the driving force for the combined charge, orbital and structural transition, where the structural transition is a secondary effect induced by the electronic order.

\begin{acknowledgments}
This research was carried out on the XPP Instrument at at the Linac Coherent Light Source (LCLS) at the SLAC National Accelerator Laboratory. LCLS is an Office of Science User Facility operated for the US Department of Energy Office of Science by Stanford University. Use of the Linac Coherent Light Source (LCLS), SLAC National Accelerator Laboratory, is supported by the U.S. Department of Energy, Office of Science, Office of Basic Energy Sciences under Contract No. DE-AC02-76SF00515. Static resonant x-ray diffraction experiments were performed at the X04SA Material Science beamline at the Swiss Light Source, Paul Scherrer Institut, Villigen, Switzerland. This work was supported by the NCCR Molecular Ultrafast Science and Technology (NCCR MUST), a research instrument of the Swiss National Science Foundation (SNSF).

\end{acknowledgments}

%

\end{document}